\documentstyle[amssymb,epsfig,12pt]{article}
\newcommand{\be}{\begin{equation}}
\newcommand{\bea}{\begin{equationarray}}
\newcommand{\ee}{\end{equation}}
\newcommand{\ea}{\end{equationarray}}
\newcommand{\ow}{\overline{W}}

\begin{document}

\title{\bf Wealth condensation in a 
simple model of economy}

\vskip 3 true cm

\author{Jean-Philippe Bouchaud$^{1,2}$, Marc M\'ezard$^3$}

\date{\it $^1$ Service de Physique de l'\'Etat Condens\'e,
 Centre d'\'etudes de Saclay, \\ Orme des Merisiers, 
91191 Gif-sur-Yvette Cedex, France \\
$^2$ Science \& Finance, 109-111 rue Victor Hugo,
92532
Levallois {\sc cedex}, France;\\ http://www.science-finance.fr\\ 
$^3$ Laboratoire de Physique Th\'eorique de l'Ecole Normale Sup\'erieure
\footnote{UMR 8548:  Unit\'e Mixte du Centre National de la Recherche
Scientifique, et de
l'\'Ecole Normale Sup\'erieure. } , \\
24 rue
 Lhomond, 75231 Paris Cedex 05, France }


\maketitle

\begin{abstract}
We introduce a simple model of economy, where the time
evolution is described by an equation capturing both exchange
between individuals and random speculative trading, in such a way that
 the fundamental symmetry of the economy under an arbitrary change of monetary units
is insured. We investigate a mean-field limit of this equation and show that the distribution
 of wealth is of the Pareto (power-law) type. The Pareto behaviour of the tails of this distribution
 appears to be robust for finite range models, as 
shown using both a mapping to the random `directed polymer' problem, as well as numerical 
simulations. In this context, a transition
between an economy dominated by a few individuals from a
 situation
where the wealth is more evenly spread out, is found. An interesting outcome is that the distribution 
of wealth tends to be very broadly distributed when exchanges are limited, either in
 amplitude or topologically. Favoring exchanges (and, less surprisingly, increasing taxes) 
seems to be an efficient way to reduce inequalities.

\end{abstract}

\vskip 0.5cm
 LPTENS preprint 00/06

\vskip 0.5cm
Electronic addresses : 
bouchaud@spec.saclay.cea.fr
mezard@physique.ens.fr

\vfill
\eject

It is a well known fact that the individual wealth is a very broadly distributed
quantity among the population. Even in developed countries, it is common that
$90\%$ of the total wealth is owned by only $5\%$ of the population. The distribution of
 wealth is often described by `Pareto'-tails, which decay as a
power-law for large wealths \cite{Pareto,Mandelbrot,Others}:
\be
{\cal P}_> (W) \sim  \left(\frac{W_0}{W}\right)^\mu,
\ee
where ${\cal P}_> (W)$ is the probability to find an 
agent with wealth greater than $W$, and $\mu$ is a certain exponent,
of order $1$ both for 
individual wealth or company sizes (see however \cite{Stanley}).

Here, we want to discuss the appearance of such Pareto tails on the basis 
of a very general model for the growth and redistribution of wealth, that we
discuss in some simple limits. We relate this model to the so-called `directed 
polymer' problem in the physics literature \cite{HHZ}, for which a large number of results
are known, that we translate into the present economical framework. We discuss the influence of
 simple parameters, such as the connectivity of the exchange
network, the role of income 
or capital
taxes and of state redistribution of wealth, on
the value of the exponent $\mu$. One of the most interesting output of such a model is the generic
 existence of a {\it phase transition}, separating a phase
where the total wealth of a very large population is concentrated in the hands
of a finite number of individuals (corresponding, as will be discussed below,
to the case $\mu < 1$), from a phase where it is shared by a finite fraction
of the population. 

The basic idea of our model is to write a stochastic dynamical equation for the
wealth $W_i(t)$ of the $i^{th}$ agent at time $t$, that takes into account the
exchange of wealth between individuals through trading, and is consistent with
the basic symmetry of the problem under a change of monetary units. Since the
unit of money is arbitrary, one indeed expects that the equation governing 
the evolution of wealth should be invariant when all $W_i$'s are multiplied by 
a common (arbitrary) factor. The evolution equation that we consider is therefore the following:
\be
\frac{d W_i}{d t} = \eta_i(t) W_i + \sum_{j (\neq i)} J_{ij} W_j
-  \sum_{j (\neq i)} J_{ji} W_i \ , \label{fund}
\ee 
where $\eta_i(t)$ is a gaussian
 random variable of mean $m$ and variance $2\sigma^2$, which describes the spontaneous growth
or decrease of wealth due to investment in stock markets, housing, etc.,
while the terms involving the (assymmetric) matrix $J_{ij}$ describe the
amount of wealth that agent $j$ spends buying the production of agent $i$ (and
vice-versa). It is indeed reasonable to think that the amount of money
earned or spent
by each
economical agent
 is proportional to its wealth.
 This makes equation (\ref{fund}) invariant under the scale transformation $W_i \to \lambda W_i$.
Technically the above stochastic differential equation is interpreted in the Stratonovich
sense \cite{Strato}. 

The simplest model one can think of is the case where all agents exchange with 
all others at the same rate, i.e $J_{ij}\equiv J/N$ for all $i \neq j$. Here,
$N$ is the total number of agents, and the scaling $J/N$ is needed to make the limit $N \to \infty$ well defined. In this case, the equation for $W_i(t)$
becomes:
\be
\frac{d W_i}{d t} = \eta_i(t) W_i + J (\ow -W_i), \label{mf}
\ee 
where $\ow=N^{-1}\sum_i W_i$ is the average overall wealth.
 This is a `mean-field' model since all agents feel the very same influence of their environment.
By formally integrating this linear equation and summing over $i$,
one finds 
that the average wealth becomes deterministic in the 
limit $N \to \infty$:
\be
 \ow(t)
=\ow(0)\exp((m +\sigma^2)t). \label{ov}
\ee 
It is useful to rewrite eq. (\ref{mf}) in terms of the normalised wealths $w_i
\equiv W_i/\ow$. This leads to:

\be
\frac{d w_i}{d t} = (\eta_i(t)-m-\sigma^2) w_i + J (1 -w_i), \label{mf2}
\ee
to which one can associate the following Fokker-Planck equation for the evolution
of the density of wealth $P(w,t)$:

\be
\frac{\partial P}{\partial t}=  \frac{\partial [J(w-1)+{\sigma^2}w]P}{\partial w}
+ {\sigma^2} 
\frac{\partial}{\partial w}\left[w \frac{\partial wP}{\partial w}\right].\label{fp}
\ee
The equilibrium, long time solution of this equation is easily shown to be:
\be
P_{eq}(w) = {\cal Z} \frac{\exp-\frac{\mu-1}{w}}{w^{1+\mu}} \qquad \mu \equiv 1 + \frac{J}{\sigma^2},
\label{Peq}
\ee
where ${\cal Z}=(\mu-1)^\mu/\Gamma[\mu]$ is the normalisation factor.
One can check that $\langle w \rangle \equiv 1$, as it should.

Therefore, one finds in this model that the distribution of wealth
exhibits a Pareto power-law tail for large $w$'s.  In agreement with intuition,
the exponent $\mu$  grows
(corresponding to a narrower distribution), when exchange 
between agents is more active 
(i.e. when $J$ increases), and also when the success in individual investment
strategies is more narrowly distributed (i.e. when $\sigma^2$ decreases).

One can actually also define the above model in discrete time, by writing:
\be
W_i(t+\tau)=\left[J\tau \ow + (1-J\tau) W_i \right]e^{-V(i,t)}
\ee
where $V$ is an arbitrary random variable of mean $m\tau$ and variance $2\sigma^2 \tau$,
and $J \tau < 1$.
In this setting, this amounts to study the so-called Kesten variable \cite{Kesten}
for which the asymptotic distribution again has a power-law tail, with an exponent $\mu$ 
found to be the solution of:
\be
(1-J\tau)^\mu \langle e^{-\mu V} \rangle =  \langle e^{-V} \rangle^\mu.
\ee
Therefore, this model leads to power-law tails for a very large class of distributions
of $V$, such that the solution of the above equation is non trivial (that is if the
distribution of $V$ decays at least as fast as an exponential). Is is easy to check that $\mu$
is always greater than one and tends to $\mu=1+J/\sigma^2$ in the limit $\tau \to 0$. Let us notice
 that a somewhat similar discrete model was studied in \cite{solom} in the context of a generalized Lotka-Volterra equation. However that model has an additional term (the 
origin of which is unclear in an economic context) which breaks the symmetry under wealth rescaling, and as a consequence the Pareto tail is truncated for large wealths.

\begin{figure}
\centerline{\hbox{\epsfig{figure=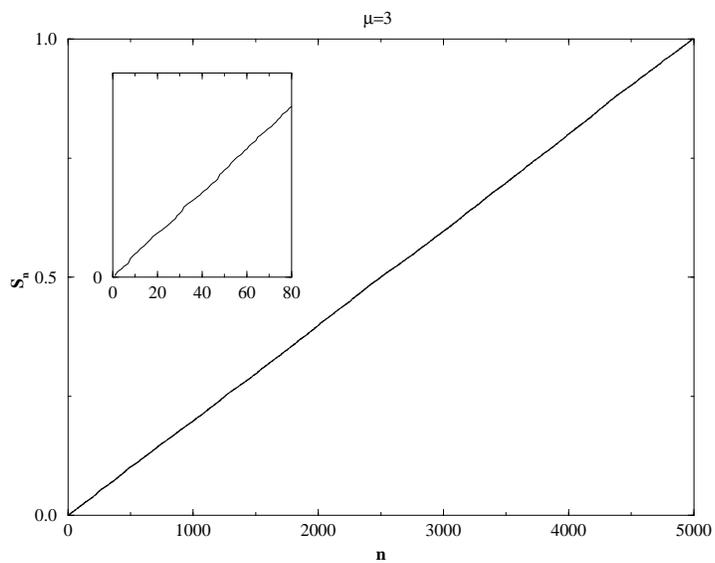,width=8cm,angle=270}}}
\vskip 0.8cm
\caption{Fraction of total wealth $S_n$ owned by the first $n$ agents, plotted versus $n$,
in a population of 5000 agents.
 The wealths have been drawn at random using a distribution with a Pareto tail exponent 
$\mu=3$. Inset: detail of the first 80 agents. One sees that $S_n$ grows 
linearly with $n$, with rather small fluctuations around the average slope $1/N$.}
\label{fig2}
\end{figure}

In this model, the exponent $\mu$ is always found to be larger than one. 
In such a regime,  if one plots the partial wealth $S_n=\sum_{i=1}^n w_i$ as a function 
of $n$, one finds an approximate straight line of slope $1/N$, with rather
small fluctuations (see Fig. 1). This means that the wealth is not too unevenly
distributed within the population. On the other hand,  the situation 
when $\mu < 1$, which we shall encounter below in some more realistic models,
 corresponds to a radically different situation
(see Fig. 2). In this case, the partial wealth $S_n$ has, for
large $N$, a devil staircase structure, with a few individuals getting hold
of a finite fraction of the total wealth. A quantitative way to measure this
`wealth condensation' is to consider the so-called inverse participation ratio $Y_2$
defined as: 
\be
Y_2 = \sum_{i=1}^N w_i^2.
\ee
If all the $w_i$'s are of order $1/N$ then $Y_2 \sim 1/N$ and tends to zero
for large $N$. On the other hand, if at least one $w_i$ remains finite when $N \to \infty$,
 then $Y_2$ will also be finite. The average value of $Y_2$ can easily be computed and is
 given by: $\langle Y_2 \rangle=1-\mu$ for $\mu < 1$
and zero for all $\mu > 1$ \cite{MPSTV,Derrida,Gumbel}. $\langle Y_2 \rangle$ is therefore a convenient
order parameter which quantifies the degree of wealth condensation.

\begin{figure}
\centerline{\hbox{\epsfig{figure=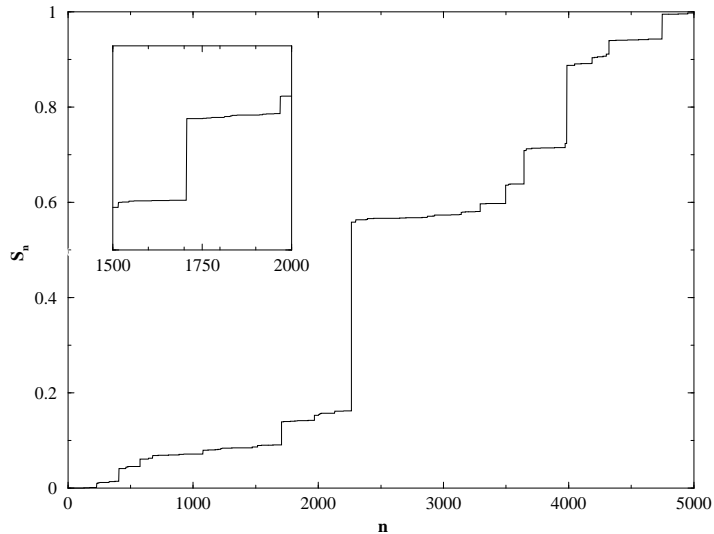,width=8cm,angle=270}}}
\vskip 0.8cm
\caption{Fraction of total wealth owned by the first $n$ agents, plotted versus $n$,
in a population of 5000 agents.
 The wealths have been drawn at random using
a distribution with a now Pareto tail exponent 
$\mu=.5 < 1$. Inset: Zoom on finer details of the curve. One clearly sees that
the curve is a `devil' staircase on all scales, with a strong dominance of 
a few individuals.}
\label{fig1}
\end{figure}

It is interesting to discuss several extensions of the above model. First, one 
can easily include, within this framework, the effect of taxes.
 Income tax 
means that a certain fraction $\phi_I$ of the income $dW_i/dt$ is taken away from
agent $i$. Therefore, there is a term $-\phi_I dW_i/dt$ appearing in the right-hand side of Eq. 
(\ref{fund}). Capital tax means that there is a fraction
$\phi_C$ of the wealth which is substracted per unit time from the wealth balance, Eq. (\ref{fund}).
 If a fraction $f_I$ of the income tax and $f_C$ of the capital 
tax are evenly redistributed to all, then this translates into a term $+f_I \phi_I d\ow/dt +
 f_C \phi_C \ow$ in the right-hand side of the wealth balance, which now reads:
\be
\frac{d W_i}{d t} = \eta_i(t) W_i + J (\ow -W_i) -\phi_I\frac{dW_i}{dt} -\phi_C W_i 
+f_I \phi_I \frac{d\ow}{dt} +
 f_C \phi_C \ow
\ee  
All these terms can be treated
 exactly within the above mean-field 
model allowing for a detailed discussion of their respective roles.
The rate of exponential growth of the average wealth $\ow(t)$ becomes equal to:
\be
\gamma \equiv \frac{m+\sigma^2/(1+\phi_I)-\phi_C (1-f_C)}{1+\phi_I (1-f_I)} .
\ee
The Pareto tail  exponent $\mu$  is now given by:
\be
\mu-1 = \frac{J(1+\phi_I)}{\sigma^2}  +\frac{1+\phi_I}{\sigma^2 (1+\phi_I (1-f_I))}
\left[ \phi_I f_I (m+\frac{\sigma^2}{1+\phi_I}) + \phi_C (f_C+\phi_I (f_C-f_I)) \right].
\ee
This equation is quite interesting. It shows that income taxes tend to reduce 
the inequalities of wealth (i.e., lead to an increase of $\mu$), even more so if
part of this tax is redistributed. On the other hand, quite surprisingly, capital 
tax, if
used simultaneously to income tax and not redistributed, leads to a {\it decrease} of $\mu$, i.e. to a wider
distribution of wealth. Only if a fraction $f_C > f_I \phi_I/(1+\phi_I)$ is 
redistributed will the capital tax be a truly social tax. Note that in the
above equation, we have implicitly assumed that the growth rate $\gamma$ is positive.
In this case, one can check that $\mu$ is always greater than $ 1+ (J+\phi_C f_C)(1+\phi_I)/\sigma^2$,
which is larger than one.

Another point worth discussing is the relaxation time associated to the Fokker-Planck
 equation (\ref{fp}). By changing variables as $w=\xi^{-2}$ 
and $P(w)=\xi^3 Q(\xi)$,
 one can map the above Fokker-Plank equation to the
one studied in \cite{times}, which one can solve exactly. For large time differences $T$,
 one finds that the correlation function of the $w$'s behaves as:
\be
\langle w(t+T) w(t) \rangle - \langle w(t) \rangle^2 \propto \exp(-(\mu-1)\sigma^2 T)\qquad \mu > 2
\ee 
and 
\be
\langle w(t+T) w(t) \rangle - \langle w(t) \rangle^2 \propto \frac{1}{(\sigma^2 T)^{3/2}}
\exp(-\mu^2 \sigma^2 T/4)\qquad \mu < 2
\ee 
This shows that the relaxation time is, for $\mu < 2$, given by $4/\mu^2\sigma^2$.
 Therefore, rich people become poor (and vice versa) on a finite time scale in this model.
 A reasonable order of magnitude for $\sigma$
is $10 \%$ per $\sqrt{\hbox{year}}$. In order to get $\mu-1 \sim 1$, one therefore has to
choose $J \sim 0.01$ per year, i.e. $1\%$ of the total wealth of an individual is
used in exchanges. [This $J$  value looks rather small, but in fact we shall see below that a more realistic (non-mean field model) allows to increase $J$ while keeeping $\mu$ 
fixed]. In this case, the relaxation time in this model is of the order of 
$100$ years.

Let us now escape from the mean-field model considered above and describe more realistic situations,
 where the number of economic neighbours to a given individual
is finite. We will first assume that the matrix $J_{ij}$ is still symmetrical, and is either equal
 to $J$ (if $i$ and $j$ trade), or equal to $0$. A reasonable first assumption is that the graph 
describing the connectivity of the population is
completely random, i.e. that two points are neighbours with probability $c/N$ and 
disconnected with probability $1-c/N$. In such a graph, the average number of 
neighbours is equal to $c$.  We  thus scale $\hat J=J/c$ in order to compare results with various
 connectivities (and insure a smooth large connectivity limit). We have performed some numerical simulations of Eq. 
(\ref{fund}) for $c=4$ and have found that the wealth distribution still has a 
power-law tail, with an exponent $\mu$ which only depends on the ratio $J/\sigma^2$.
This is expected since a rescaling of time by a factor $\alpha$ can be absorbed by changing
$J$ into $\alpha J$ and $\sigma$ into $\sqrt{\alpha} \sigma$; therefore, long time (equilibrium)
 properties can only depend on the ratio $J/\sigma^2$. As shown in Fig. 3, the exponent $\mu$ can
 now be smaller than one for sufficiently small values of $J/\sigma^2$. In this model, one
therefore expects wealth condensation when the exchange rate is too small.
Note that we have also computed numerically the quantity $\langle Y_2 \rangle$ and found
very good agreement with the theoretical value $1-\mu$ determined from the slope
of the histogram of the $w_i$'s.

\begin{figure}
\centerline{\hbox{\epsfig{figure=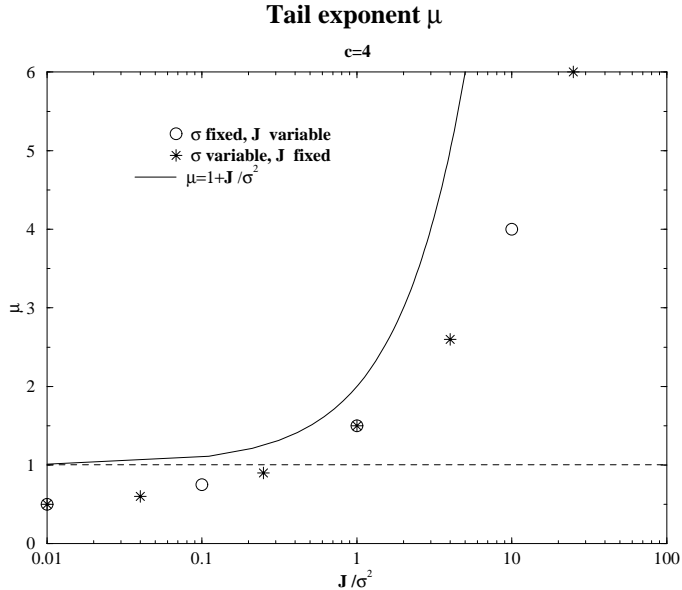,width=8cm,angle=270}}}
\vskip 0.8cm
\caption{Plot of the numerical values of $\mu$ for the model on a random graph with connectivity $c=4$,
 as a function of $J/\sigma^2$ -- we have indeed checked that this scaling holds for our numerical discretization of Eq. (\protect\ref{fund}). The plain line corresponds to the mean-field prediction $\mu=1+J/\sigma^2$.
 For $c=4$, we find that the condensation transition takes place for $J/\sigma^2 \sim 0.3$.}
\label{fig3}
\end{figure}
 
 From the physical point of view,
the class of models which we consider here belong to the general family of
directed polymers in random media. The two cases we have considered so far
correspond respectively to a polymer on a fully connected lattice, and
a polymer on a random lattice.
A variant of this model can be solved exactly using the method of Derrida and Spohn \cite{DS} for the
 so-called directed polymer problem on a tree. In this variant one assumes that at each time step $\tau$
the connectivity matrix is completely changed and chosen anew using the same probabilities as
above. Each agent $i$ chooses at random exactly $c$ new neighbours $\ell(i,t)$, the wealth evolution equation becomes
\be
W_i(t+\tau)=\left[{J\tau \over c} \sum_{\ell=1}^c W_{\ell(i,t)}+(1-J\tau) W_i(t)\right] e^{-V(i,t)}
\ee
where $V$ is a gaussian random variable of mean zero and variance $2\sigma^2 \tau$. One can 
then write a closed equation for the evolution of the wealth distribution \cite{DS}.
 In this case, the wealth condensation
phenomenon takes place whenever $\sigma^2 \tau +J \tau \ln(J \tau/c)+(1-J \tau ) \ln(1-J \tau) > 0$.
For $J \tau \ll 1$ the transition occurs for $\sigma^2 = \sigma_c^2 = J (1+ \ln(c/J\tau))$.

For $\sigma > \sigma_c$, one finds that $\mu$ is given by: 
\be
\mu \simeq \frac{\ln\left(\frac{c}{\sigma^2\tau}\right)}{\ln(c/J\tau)}
\ee
and is less than one, signalling the onset of a phase where wealth is condensed on a finite number of individuals. This precisely corresponds to the glassy phase in the directed polymer language.  The above formula shows that $\mu$ depends only weakly on $\sigma$ or $J$, in qualitative agreement with our numerical result for the continuous time model (see Fig. 3). Note that in the limit $c \to \infty$,  $\sigma_c \to \infty$ and the glassy phase disappears,
in agreement with the results above, obtained directly on the mean-field model. Note also that
in the limit $\tau \to 0$, where the reshuffling of the neighbours becomes very fast, wealth
diffusion within the population becomes extremely efficient and, as expected, the transition again disappears. Finally, in the simple case where $J\tau=1$
(each agent trading all of his wealth at each time step), the critical value is $\sigma_c^2\tau= \ln c$ and the exponent $\mu$ in the condensed phase is simply $\mu=\sigma_c/\sigma$, and  $\mu=\sigma^2/\sigma_c^2$
for $\mu > 1$ (see \cite{DS}).

Let us note, {\it en passant}, that the model considered by Derrida and Spohn has another
interesting interpretation if the $W_i$'s describe the wealth of companies. The growth of a company takes place either from internal growth (leading to a term $\eta_i(t) W_i$ much as
above), but also from merging with another company. If the merging process between two
companies is completely random and takes place at a rate $\lambda$ per unit time, then the
model is exactly the same as the one considered in Section 3 of \cite{DS} (see in particular their Eq. (3.2)).

Although not very realistic, one could also think that the individuals
are located on the nodes of a d-dimensional
hypercubic  lattice, trading with their neighbours up to a finite distance. In this case, one
knows that for $d > 2$ there exists again a phase transition between 
a `social' economy where $\mu > 1$ and a rich dominated phase $\mu < 1$. On the other hand, for $d \leq 2$,
 and for large populations, one
is always in the extreme case where $\mu \to 0$ at large times. In
the case $d=1$, i.e. operators organized along a chain-like structure,
one can actually compute exactly the distribution of wealth by transposing the results of 
\cite{FHH}. One finds for example that the ratio of the maximum wealth to the typical (e.g. median)
wealth behaves as
 $\exp \sqrt{N}$, where $N$ is the size of the population, 
instead of $N^{1/\mu}$ in the case of a Pareto distribution with $\mu > 0$. The conclusion of the above
 results is that the distribution 
of wealth tends to be very broadly distributed when exchanges are limited, either
in amplitude (i.e. $J$ too small compared to $\sigma^2$) or topologically (as in the above chain structure).
 Favoring exchanges (in particular with distant neighbours) seems to be an efficient way to reduce inequalities.

Let us now discuss in a cursory way the extension of this model to the
case where the matrix $J_{ij}$ has a non trivial structure. One can always
write:
\be
J_{ij} = D_{ij} \exp -\frac{F_{ij}}{2} \qquad
J_{ji} = D_{ij} \exp +\frac{F_{ij}}{2},
\ee
where $D_{ij}$ is a symmetric matrix describing the frequency of trading between
$i$ and $j$. $F_{ij}$ is a local bias: it describes 
by how much the amount of trading from $i$ to $j$ exceeds that from $j$ to $i$.
In the absence of the speculative term $\eta_i W_i$, Eq. (\ref{fund}) is actually
a Master equation describing the random motion of a particle subject to local {\it forces} $F_{ij}$,
 where $J_{ij}$ is the hopping rate between site $j$ and site $i$. This problem has also been much
 studied \cite{BG}. One can in general decompose the force $F_{ij}$ into a potential part $U_i-U_j$ 
and a non potential part.
For a purely potential problem, the stationary solution of Eq. (\ref{fund})
with $\eta_i \equiv 0$ is the well known Bolzmann weight:
\be
W_{i,eq}= \frac{1}{Z} \exp(-U_i) \qquad Z =\sum_{i=1}^N  \exp(-U_i). 
\ee
The statistics of the $W_i$ therefore reflects that of the potential $U_i$; in particular, large
 wealths correspond to deep potential wells. Pareto tails correspond to the case where the extreme
 values of the potential obey the
Gumbel distribution, which decays exponentially for large (negative) potentials
\cite{Gumbel}.

The general case where $\eta_i$ is non zero and/or $F_{ij}$ contains a non potential part is largely
 unknown, and worth investigating. A classification of the cases where the Pareto tails survive the
 introduction of a non trivial bias field $F_{ij}$ would be very interesting. Partial results in the
context of population dynamics have been obtained recently in \cite{Nelson}. The case where the $i$'s are on
 the nodes of a $d$ dimensional lattice should be amenable
to a renormalisation group analysis along the lines of \cite{Fisher,Luck}, with interesting 
results for $d \leq 2$. Work in this direction is underway \cite{ustocome}.

In conclusion, we have discussed a very simple model of economy, where the time
evolution is described by an equation capturing, at the simplest level, exchange
between individuals and random speculative trading in such a way that the fundamental
 symmetry of the economy under an arbitrary change of monetary units
is obeyed. Although our model is not intended to be fully realistic, the family of equations given by Eq. (\ref{fund}) is 
extremely rich, and leads to interesting generic predictions. We have investigated in details a mean-field limit of this equation and showed
 that the distribution of wealth is of the Pareto type. The Pareto behaviour of the tails of this
 distribution appears to be robust for more general connectivity matrices, as a mapping to the
 directed polymer problem shows. In this context, a transition
between an economy governed by a few individuals from a  situation
where the wealth is more evenly spread out, is found. The important conclusion of the above model
 is that the distribution 
of wealth tends to be very broadly distributed when exchanges are limited. Favoring exchanges (and,
 less surprisingly, increasing taxes) seems to be an efficient way to reduce inequalities.

\vskip 1cm

Acknowledgments: We want to thank D.S. Fisher, I. Giardina and D. Nelson for interesting 
discussions. MM thanks the SPhT (CEA-Saclay) for its hospitality.

\end{document}